\documentclass[twocolumn,amsmath,amssymb]{revtex4}
\usepackage{subfigure}
\usepackage{here}
\usepackage{float}
\usepackage{wrapfig}
\usepackage{graphicx}
\usepackage{dcolumn}
\usepackage{bm}
\usepackage{amsmath}
\usepackage{amsfonts}
\usepackage{amssymb}
\usepackage{color}
\usepackage{mathrsfs}
\makeatletter

\begin{document}
\title{Mechanics of a snap-fit}
\author{Keisuke Yoshida and Hirofumi Wada} 
\affiliation{Department of Physical Sciences, Ritsumeikan University, Kusatsu, Shiga 525-8577, Japan}

\begin{abstract}
Snap-fits are versatile mechanical designs in industrial products, which enable the repeated assembling and disassembling of two solid parts. This important property is attributed to a fine balance between geometry, friction, and bending elasticity. In the present study, we combine theory, simulation, and experiment to reveal the fundamental physical principles of snap-fit functions in the simplest possible setup consisting of a rigid cylinder and a thin elastic shell. We construct a phase diagram using geometric parameters and identify four distinct mechanical phases. We develop analytical predictions based on the linear elasticity theory combined with the law of static friction and rationalize the numerical and experimental results. The study reveals how an operational asymmetry of snap-fits (i.e., easy to assemble but difficult to disassemble) emerges from an exquisite combination of geometry, elasticity, and friction and suggests optimization of the tunable functionalities for a range of mechanical designs. 
\end{abstract}
\maketitle

Assembling and disassembling two solid components is a fundamental process for functional structures in natural and manmade systems. Examples encompass different length scales ranging from ligand-receptor interactions in biochemistry~\cite{Moy-Science-1994} and plastic shell covers in industrial products~\cite{Sui-JMechDes-2000} to the docking of free-flying space vehicles~\cite{Romano-JSpaceRoc-2007}. 
 In manufacturing industries, snap-fits are typically used to join two plastic parts without gluing, constituting a simple design with necessary resilience to allow for repeated assembly and disassembly ~\cite{Bayer-Report-1996}. 
Today, snap-fits are found everywhere including the cap of a marker pen, plastic zipper bag, and toys, such as Lego blocks.
The "click" sound is a familiar snap-fit characteristic present in our daily lives. 
In most snap-fit designs, assembling requires a relatively less effort whereas disassembling is more difficult. 
Mechanical asymmetry is a central property of snap-fits for industrial use and emerges from an interplay between flexibility, frictional interactions, and the geometric structure of the snap-fit parts. 
However, despite its familiarity and prevalent use in daily life, fundamental aspects of snap-fit mechanics are highly unexplored, at least from the perspective of physics. 

\begin{figure}[b]
 \includegraphics[width=0.97\linewidth]{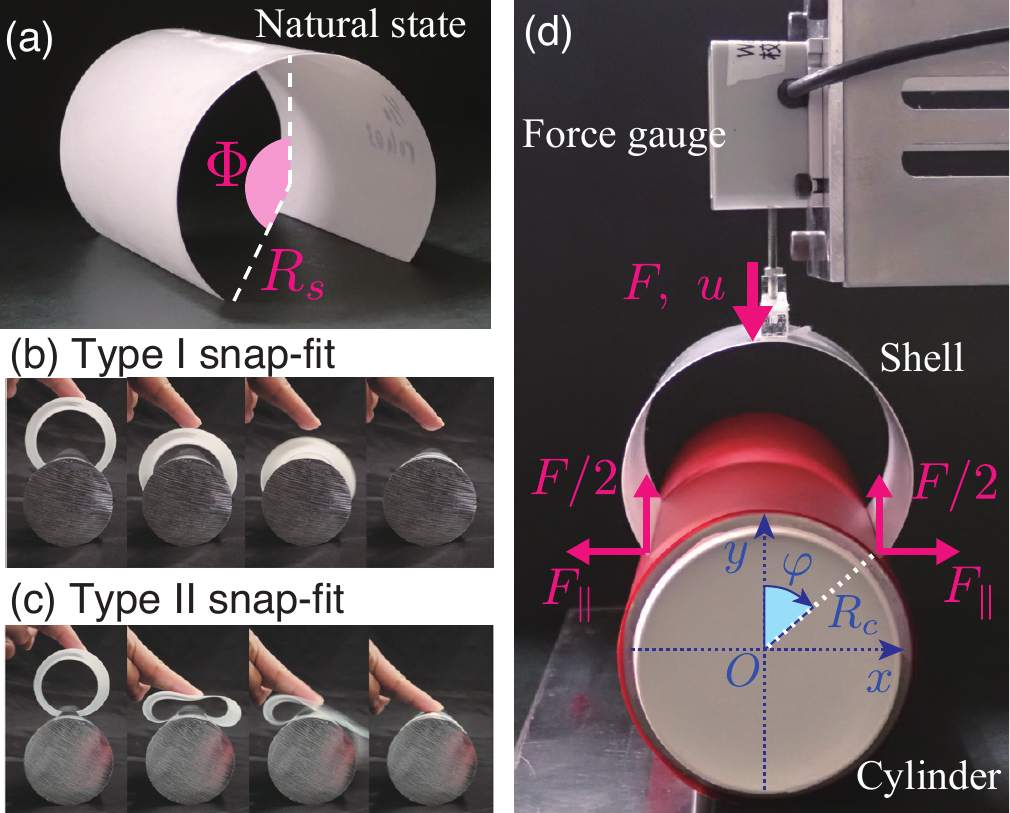}
\caption{(a) Definition of geometric parameters of the shell, i.e., radius $R_s$ and opening angle $\Phi$ in its natural configuration.
(b-c) Sequence of snapshots of a thin plastic shell undergoing Type I snap-fit, (b) and Type II snap-fit (c) process. 
(d) Schematic photographic view of our experimental system for force measurements during assembling and disassembling processes.}
\label{fig:figure1}
\end{figure}

In the study, we propose a model experimental system that can illustrate snap-fit behavior; we investigate its physical properties by combining force measurements, numerical simulations, and theoretical analysis. 
We consider a semi-cylindrical shell of radius $R_s$ and thickness $t$ [Fig.~\ref{fig:figure1} (a)], which is pushed onto a surface of a rigid cylinder with radius $R_c$ [Fig.~\ref{fig:figure1} (b) and (c)]. 
The shell either clutches the cylinder via a snap-fit or buckles on the cylindrical surface, based on their respective geometries. We integrate experimental and numerical data to construct a phase diagram in terms of the geometric parameters and rationalize the observed phase boundaries with an analysis based on linear elasticity theory combined with a dry friction law~\cite{Benson-JAppMech-1981, Benson-JAppMech-1982, He-ArchApplMech-1997, Kopp-TextResJ-2000, Liu-IntJSolStruct-2013, Chateau-EurJMechA-1991, Roman-JMechPhysSolids-2002, Nguyen-Mechnique-2003}. 
The proposed model is minimal, albeit versatile, and potentially scalable and can be used as a building block in the design of artificial non-reciprocal mechanical metamaterials~\cite{Rafsanjani-AdvMat-2015, Shan-AdvMat-2015, Coulais-Nature-2017}.

The problem is essentially two-dimensional as a shell deforms uniformly along the cylindrical axis, with a length of 20 mm.
Hence, we focus on the shapes of the shell cross-section, with relevant geometric parameters corresponding to the radius ratio $\alpha=R_c/R_s$ and opening angle, $\Phi$ [Fig.~\ref{fig:figure1} (a)]. 
 A shell is prepared by adding a permanent intrinsic curvature to an initially flat sheet of polystyrene ($t=0.2, 0.3$, and 0.4 mm) via thermoforming with hot water. 
 Image analysis of the cross-sectional shape in the resulting semi-cylinder confirms a sufficiently uniform radius of curvature that ranges from $R_s =25.2-29.5$ mm with various angles in the range of $\Phi=1.8-3.0$ rad. 
The bending moduli of the shells $B= 6.7\times 10^{-5}-4.3\times 10^{-4}$ ${\rm N\cdot m^2}$ for shell thickness $t=0.2-0.4$ mm are independently measured. 
The shell is sufficiently stiff such that the effects of gravity are negligible.
We cover the surface of an acrylic cylinder with radius $R_c = 30, 35$, and 40 mm, with a thin oriented polypropylene (OPP) sheet of thickness 10 $\mu$m to ensure uniform frictional interactions with the shell.
The stepping motor controls the vertical position of the top of the shell through a force gauge while the rigid cylinder is fixed onto a bottom substrate [Fig.~\ref{fig:figure1} (d)]. 
During the assembling process, the shell moves downwards at a speed of 5 mm/s until the top of the shell touches the cylindrical surface. 
After a 6 s interval, the shell then moves upwards with the same speed of 5 mm/s (i.e., the disassembly process). 
A "push-back" force $F$ exerted by the shell is measured via a force gauge attached to the top of the shell. 
A snap instability in the assembling process implies that a force curve crosses $F=0$ from positive to negative. 
In the $F>0$ region, the shell repels from the cylinder without loading, whereas the shell spontaneously clutches the cylindrical surface without any further loading in the $F<0$ region. 
The measured force $F$ is displayed in units of $B/R_s^2$.

\begin{figure*}
 \includegraphics[width=0.97\linewidth]{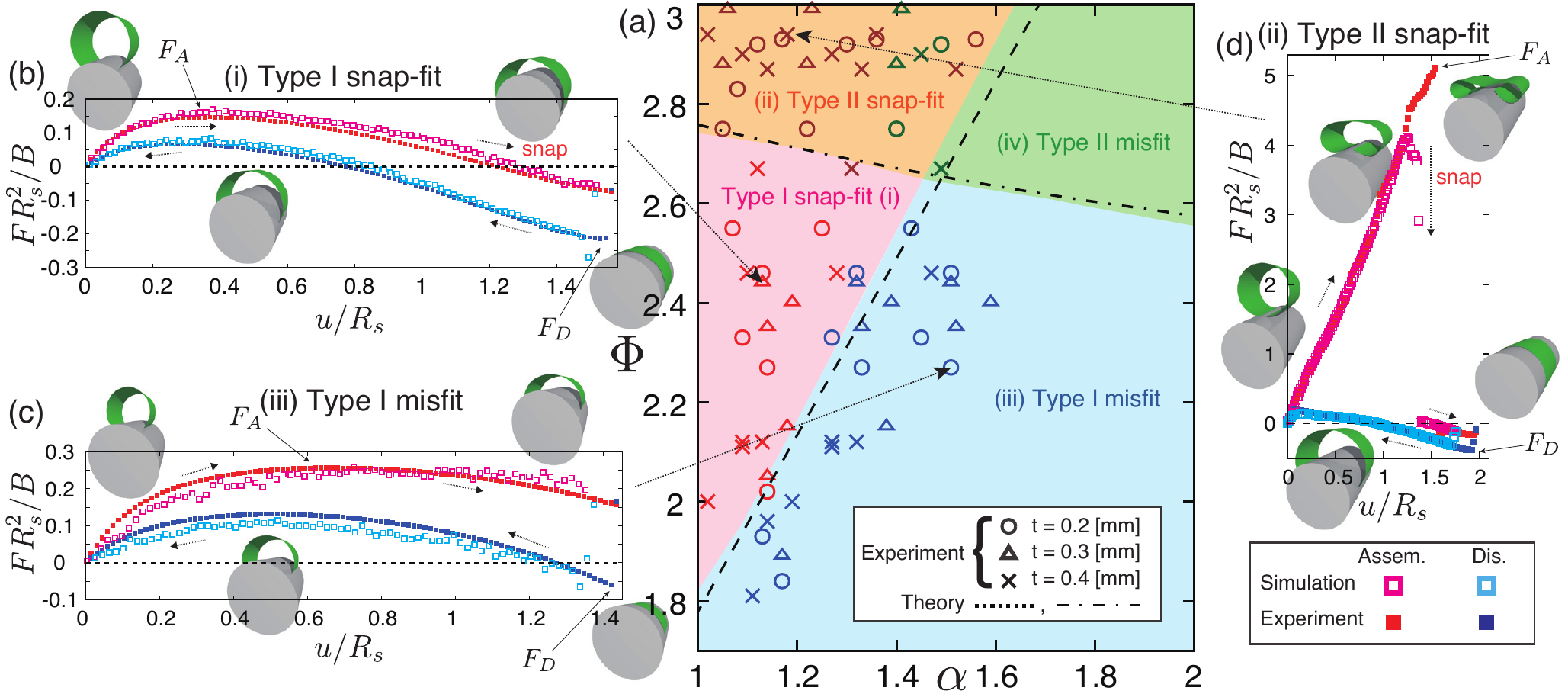}
\caption{(a) Mechanical phase diagram in the assembling process drawn on the $(\alpha, \Phi)$ parameter space. Open symbols denote data from experiments for different shell thickness $t=0.2$ mm (circle), $t=0.3$ mm (triangle), and $t=0.4$ mm (cross). 
Red and orange regions denote the snap-on domain whereas blue and green regions denote the misfit domain, which are identified by the numerical simulations performed on regularly arranged grids on the $(\alpha,\Phi)$ plane (every 0.1 for both $\alpha$ and $\Phi$).
Dashed and dash-dotted lines represent analytical curves, as explained in the main text. 
(b--d) The rescaled force of a shell $F R_s^2/B$ as a function of displacement rescaled by shell radius, $u/R_s$, during the cyclic process. 
Open red and blue symbols denote data from the assembly and disassembly experiments, respectively, whereas filled symbols represent corresponding simulations. The specific values of $(\alpha, \Phi)$ for (b), (c), and (d) are indicated in (a). 
Shell configurations inserted in (b)--(d) are reproduced from the simulations. }
\label{fig:diagram}
\end{figure*}

To complement the experimental results, we also performed numerical simulations using a discrete analog of the continuum elastica model~\cite{Sano-PRL-2017}.  
The frictional interaction between the shell and cylindrical surface is modeled based on Amontons--Coulomb's law, which states that the contact point remains stationary if the tangential force is below the critical value $\mu P$, where $P$ denotes the normal reaction and $\mu$ denotes the coefficient of static friction~\cite{Popova-Friction-2015}. 
 (Full details of the numerical method are given in Supplemental Material~\cite{SM}.) 
We compare the experimental force curves with those obtained from the simulations in Fig.~\ref{fig:diagram} (b)--(d) and an excellent agreement is realized. From this, the value of $\mu$ in the experiments is typically determined as $\mu=0.21$. 

Experimental and numerical investigations are summarized in Fig.~\ref{fig:diagram} (a). 
We identify four different phases that are broadly divided into two distinct domains, i.e., the snap-on and misfit domains.
The snap-on domain consists of two phases, referred to as (i) Type I snap-fit and (ii) Type II snap-fit phases.
The misfit domain consists of the remaining two phases, referred to as (iii) Type I misfit and (iv) Type II misfit. 
In two Type I phases [(i) and (iii)], a shell is only moderately deflected.
For snap-fit phase (i), the shell exhibits a snap instability at $F=0$ and fits the cylinder spontaneously. 
In contrast, for the misfit phase (iii), $F$ remains positive during the entire pushing process and the shell makes no snap-fit.
In both cases, the groove of the shell initially opens and the force rapidly increases [Fig.~\ref{fig:diagram} (b) and (c)]. 
Past the force maximum, it decreases as the vertical displacement $u$ increases. 
For a sufficiently deep shell of $\Phi > \Phi_{\rm sf}(\alpha)$ (explained below), the force crosses $F=0$ to snap~\cite{
Patricio-PhysicaD-1998,Holms-AdvMat-2007,Chen-EurJMechA-2011,Pandey-EPL-2014,Gomez-NatPhys-2016}. 
See Supplemental Material for Video S1~\cite{SM}.
For $\Phi < \Phi_{\rm sf}(\alpha)$, the top of the shell touches the cylindrical surface before $F$ corresponds to zero. 
If the loading is removed, the shell either pushes back or retains its instantaneous shape based on $\mu$. 
For a detailed discussion of this, see \cite{SM}.
In the Type I region, the force discontinuously assumes a substantial negative value when the shell is pulled out from the cylinder; this is because the normal reaction increases discontinuously when its vertical component changes sign as the loading switches from pushing to pulling.
The discontinuous force jump in Fig.~\ref{fig:diagram} (b)--(d) manifests the critical role of static friction on the hysteric non-reciprocal responses. 
The simulations confirmed that a force response is completely reversible in the absence of friction, $\mu=0$~\cite{SM}. 
In contrast to a common cantilever snap-fit design~\cite{Bayer-Report-1996, Ji-JMechDes-2011}, purely geometry-based asymmetry is absent. 
A key issue in the present study is that the asymmetry is due to the coupling between the geometry and the friction.

Conversely, Type II phases [(ii) and (iv)] involve high deflections.
It is noted that the maximum assembling force in (ii) is approximately ten times that in (i).
In (ii) and (iv), a shell is strongly squeezed to assume an M-shaped configuration with two ends rolled up.
At a critical compression, the two ends suddenly jump outwards such that the shell is outstretched to create a surprisingly loud snapping sound.
See Supplemental Material for Video S2~\cite{SM}.
The shell eventually "snaps-on" to the cylindrical surface in (ii), whereas the misfit is observed in (iv) as $\alpha$ is excessively high, in which the shell bounces back (or maintains its instantaneous shape) if the loading is removed. 
It is noted that the assembling force in (ii) significantly exceeds the disassembling force, contrary to expectations in industrial designs. However, the behavior is potentially interesting in terms of efficient energy-absorbing devices~\cite{Wu-AdvEngMat-2018}. 
Specifically, the diagram is overall insensitive to the modulus $B$, thickness $t$, and friction coefficient (if $0.2<\mu<0.5$), thereby indicating that it is highly geometrical.
 
To understand the qualitatively distinct behaviors, we now develop an analytical argument.
We first define a coordinate system, as shown in Fig.~\ref{fig:figure1} (d), and remark that an important insight is obtained from the simulations. 
For most parameter sets $(\alpha,\Phi,\mu)$, a shell touches a cylinder only at its two edges, thereby indicating that the contact always corresponds to point-like (or, line-like in a 3-dimensional view). 
This might be counter-intuitive because an evidently areal contact is typically observed between the shell and cylinder surface around snap-on configurations in experiments. 
However, discrete contact is a direct consequence of the moment-free boundary conditions at the shell edges combined with the mismatch of two natural curvatures, i.e., $\alpha \neq 1$~\cite{Goriely-PRL-2006}. 
Given that external forces are only applied at three discrete points on the shell, the overall vertical force balance requires that the sum of the forces must vanish irrespective of the shape of a shell, $0 = -F + 2P \cos\varphi+2Q \sin\varphi$, where $P$ and $Q$ denote the normal and tangential components of the force exerted from the surface, and the contact point of the end of the shell has an angle $\varphi$ to the vertical $(y)$ axis. See Fig.~\ref{fig:figure1} (d).
In a quasi-static process, the critical condition $Q=\mu P$ can hold, and this leads from the above force balance to the following expression: 
\begin{eqnarray}
 \frac{F}{F_{\parallel}} &=& \frac{2(1+\mu\tan\varphi)}{\tan\varphi-\mu},
 \label{eq:F-formula}
\end{eqnarray}
where $F_{\parallel}=P\sin\varphi-Q\cos\varphi$ denotes the horizontal component of the force~\cite{Mose-MecEngSci-2019}. 
A similar formula is valid for the disassembling process, with the replacement given by $\mu \rightarrow -\mu$ in Eq.~(\ref{eq:F-formula}). 
We see from Eq.~(\ref{eq:F-formula}) that the snap-fit point, $F=0$, is given by $1+\mu\tan\varphi^{\ast}=0$. 
For snap-fit bifurcation to occur, the (half) contour length of the shell, $R_s\Phi$, must exceed the arclength along the cylindrical surface, $R_c\varphi^*$, and thus we predict the following:
\begin{eqnarray}
 \Phi > \Phi_{\rm sf} \approx \alpha \left(\pi-\tan^{-1} \frac{1}{\mu} \right).
 \label{eq:snap-fit-cond}
\end{eqnarray}
As shown in Fig.~\ref{fig:diagram} (a), Eq.~(\ref{eq:snap-fit-cond}) explains the phase boundary between (i) and (iii) very well.
For $\mu \rightarrow 0$, we obtain $\Phi_{\rm sf} \rightarrow (\pi/2)\alpha$. 
A shell snaps when it crosses the "equator" of a cylinder, confirming our intuition. 

To explain the Type I and II boundaries, we need to develop an analysis based on the theory of elastica with natural curvatures~\cite{Frisch-Fay-Book, Nordgren-IntJSolStruct-1966, Chateau-EurJMechA-1991,Roman-JMechPhysSolids-2002} and Amontons' law, and obtain $P$ and $Q$ as functions of $\alpha, \Phi$, and $F$. 
If the edges of the shell are pinned as soon as the shell touches the cylindrical surface at $\varphi=\arcsin(\alpha^{-1}\sin\Phi)$, the pinned configuration is increasingly stabilized when the shell is compressed further and finally leads to a high amplitude Type II snap. 
The linear response theory given in Supplemental Material~\cite{SM} allows the ratio $Q/P$ to be independent of $F$; thus, the critical condition $\mu=Q/P$ leads to an analytical expression in terms of $\alpha, \Phi$, and $\mu$ only. 
The resulting implicit expression is numerically solved to yield $\Phi_{\rm I-II}=\Phi_{\rm I-II}(\alpha, \mu)$, which is in excellent agreement with the simulation and experiment, as shown in Fig.~\ref{fig:diagram} (a).

In an ideal snap-fit design, two solid parts are reasonably easy to assemble, while disassembling the same is more difficult, albeit not excessively. 
This type of medium asymmetry can be read in force curves in Fig.~\ref{fig:diagram} (b) and (d) as the criterion in which the magnitude of the maximum force in the disassembly process, $|F_D|$, is higher although not excessively higher than that in the assembly process, $F_A$. 
(Note that $F_D<0$ for a snap-on configuration.)

\begin{figure}
 \includegraphics[width=0.99\linewidth]{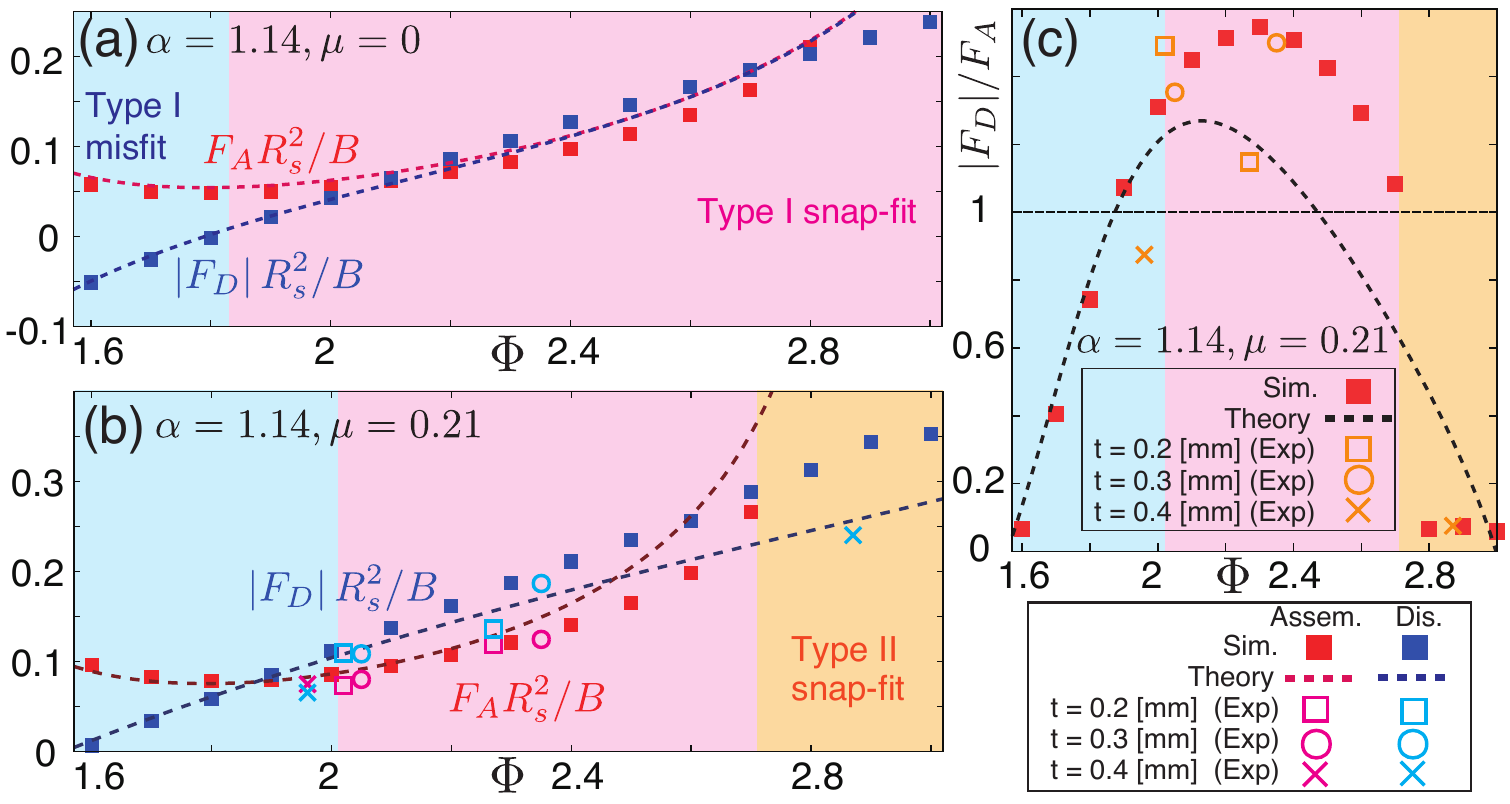}
\caption{
(a-b) Rescaled assembling and disassembling forces, $F_A R_s^2/B$ and $|F_D| R_s^2/B$, as a function of angle $\Phi$ for $\alpha=1.14$.
(a) $\mu=0$ and (b) $\mu=0.21$. (c) Ratio $|F_D|/F_A$ vs. $\Phi$ for $\mu=0.21$ from the same data as shown in (b). Open and filled symbols denote the experimental and numerical data, respectively. Dotted lines denote the theoretical predictions given in the main text. }
\label{fig:forces}
\end{figure}

 In Fig.~\ref{fig:forces} (a) and (b), $F_A$ and $|F_D|$ obtained from the experiment and simulation are plotted as functions of $\Phi$ for the friction-less $(\mu=0)$ and frictional $(\mu=0.21)$ cases, respectively. 
 In Fig.~\ref{fig:forces} (c), the data for $\mu=0.21$ are replotted in a form as $\Phi$ vs.~$|F_D|/F_A$, a metric defined as "locking ratio" in Ref.~\cite{Sui-ResEngDes-2000}. 
 Overall, a desirable condition $|F_D|/F_A>1$ is achieved for $2<\Phi<2.6$, and this highly overlaps with the Type-I snap-fit regime.
The trend is valid for other typical values of $\mu$, thereby suggesting that a relative magnitude of $|F_D|$ and $F_A$ can only be tuned with shell geometry, $\Phi$. 

We now rationalize the above findings with the linearized theory that we have developed above. 
We first present the results for the frictionless ($\mu=0$) case, for which a shell deformation is dominated by the horizontal component of the normal reaction $F_{\parallel}$, since it is applied at the free edge of the shell~\cite{Matsumoto-EPL-2018}. Conversely, the vertical compression force, $F$, deforms the shell much less than $F_{\parallel}$, even though its magnitude is not necessarily small. We can thus approximately predict $F$ by knowing $F_{\parallel}$ through Eq.~(\ref{eq:F-formula}) with $\mu=0$, i.e., $F/F_{\parallel}=2/\tan\varphi$, for a given configuration of the shell without $F$. This simplification allows us to obtain compact analytical expressions for $F_A$ and $F_D$, as shown below.

For $\mu=0$, we consider a small deformation of a naturally curved elastica subject to outgoing horizontal forces $F_{\parallel}$ applied at the two edges. 
By expanding relevant equations in terms of $F_{\parallel} R_s^2/B$ up to the first order and imposing the inextensibility constraint, we obtain 
a linear relation $F_{\parallel}R_s^2/B = K(\Phi) \Delta/R_s$, where $\Delta$ denotes the horizontal displacement and 
\begin{eqnarray}
 K(\Phi) &=& \left[\frac{1}{2}\Phi- \cos\Phi \left(\frac{3}{2}\sin\Phi -\Phi\cos\Phi\right)\right]^{-1}
 \label{eq:K}
\end{eqnarray}
denotes the $\Phi$-dependent effective spring constant.  
 (For details, see our Supplemental Material~\cite{SM}.) 
We combine this with Eq.~(\ref{eq:F-formula}) to obtain an analytic expression for $F$ in terms of $\Delta$, 
where $\varphi$ is related to $\Delta$ and $\Phi$ via $\alpha\sin\varphi=\sin\Phi+\Delta/R_s$. 
For $\mu=0$, it takes a particularly compact form given by $FR_s^2/B = 2 K(\Phi) \cot\varphi (\alpha\sin\varphi-\sin\Phi)$.
We maximize this with respect to $\varphi$ and obtain the following expression
\begin{eqnarray}
 \frac{F_A R_s^2}{B} &=& 2\alpha K(\Phi) \left[1-\left(\frac{\sin\Phi}{\alpha}\right)^{2/3}\right]^{3/2},
 \label{eq:F_a-mu=0}
\end{eqnarray}
which is in excellent agreement with the simulation data [Fig.~\ref{fig:forces} (a)].
The disassembling force $|F_D|$ is evaluated for a shell that is slightly deformed from its snap-on configuration, for which $\varphi\approx \Phi/\alpha$.
A similar analysis to $F_A$ then leads to 
\begin{equation}
 \frac{F_D R_s^2}{B}=2\alpha K(\Phi)\frac{\sin(\Phi/\alpha)-\alpha^{-1}\sin\Phi}{\tan(\Phi/\alpha)} 
 \label{eq:F_d}
\end{equation}
Again, Eq.~(\ref{eq:F_d}) is observed to agree well with our simulation in Fig.~\ref{fig:forces} (a).

A frictional case $(\mu>0)$ is analyzed similarly, although it is mathematically more involved, because we need to explicitly account for the effects of the vertical force $F$ on the elastica shape. 
The linearized theory that has been developed to explain the Type I and II boundaries, $\Phi_{\rm I-II}(\alpha,\Phi)$, is reproduced here. 
After some straightforward calculations, we can obtain similar formulae to Eqs.~(\ref{eq:F_a-mu=0}) and (\ref{eq:F_d}), which are given in Supplemental Material~\cite{SM}. 
The analytical predictions for $F_A$ and $F_D$ for $\mu=0.21$ are compared with the numerical and experimental data in Fig.~\ref{fig:forces} (b) and (c), 
showing a quantitative agreement for $\Phi < 2\pi/3$. 
The agreement is only qualitative for larger $\Phi$, for which the linearized theory becomes systematically inaccurate as the deflections of the shell become larger for the increase in $\Phi$.

Figure~\ref{fig:forces} (b) suggests that $F_A$ increases with $\Phi$ and diverges as $\Phi\rightarrow \Phi_{\rm I-II}$, at which the system enters the Type II phase. 
Figure~\ref{fig:forces} (b) also shows that the disassembling force $|F_D|$ increases significantly with $\Phi$.
Such a "locking" phenomenon occurs as the external pulling force increases the normal reaction, increasing the tangential component of the friction force and thus the external force required for disassembling.  
The friction-mediated self-stiffing mechanism exists for structures with curved geometry and is more prominent for a deeper shell.

Interestingly, the snap-fit behavior that is favorable for industrial applications, i.e., the Type I snap-fit, is only achieved for a limited range of geometric designs.
To expand the design space for snap-fits, this formulation should be generalized to account for other mechanical aspects, such as shell stretch and cylinder deformation, which are likely to be important for thicker shells and soft constituent materials, e.g., elastomers.
In real experiments, air-flow induced effects, such as a negative pressure via the suction of air during snapping, can alter frictional interactions.
Therefore, the estimated value, $\mu=0.21$, can be interpreted as an effective value that accounts for the aforementioned physical effects. 
In association with this, the validity of Amontons' law and the possibilities of other friction laws may have to be examined based on the systems under consideration.

To the best of the authors' knowledge, this is the first detailed study of snap-fit mechanics in the context of the physics of thin structures~\cite{Audoly-Book, Holms-Review-2019}. 
Our study, which combines experiment, numerical simulation, and linear elasticity theory, reveals a quantitative design space for snap-fits and illustrates how an exquisite combination of geometry, elasticity, and friction leads to an emergent mechanical asymmetry between assembling and disassembling processes. 
The proposed model is potentially scalable, and the clarified route for such non-reciprocal force responses can inspire a new class of energy-absorbing metameterials~\cite{Rafsanjani-AdvMat-2015, Shan-AdvMat-2015, Coulais-Nature-2017}. 
The study is also potentially insightful to envision a future snap-fit design suitable for sustainable materials, which can ultimately contribute to reducing plastic waste.  

\begin{acknowledgments}
H.W. thanks the Isaac Newton Institute for Mathematical Sciences for support and hospitality during the workshop "GFS follow on: Mathematics of form in active and inactive media", where the discussions with the participants were useful to complete the study. 
We acknowledge financial support from JSPS KAKENHI (Grants No.~18K13519 and No.~18K18741, to H.W.) and the Sasakawa Scientific Research Grant from The Japan Science Society (to K. Y.).
\end{acknowledgments}


\end{document}